\documentstyle[12pt]{article}
\begin{document}

\begin{center}

{\large\bf (Anti-)self-dual homogeneous
 vacuum gluon field as an origin of confinement and
$SU_L(N_F)\times SU_R(N_F)$
symmetry breaking in QCD}\\

\vspace*{.5cm}

{\bf G.V.~Efimov}\\

{\footnotesize\it Bogoliubov Laboratory of Theoretical Physics,
Joint Institute  for Nuclear Research, 141980  Dubna,
Moscow Region, Russia}\\

{and}\\

{\bf S.N.~Nedelko}\\

{\footnotesize\it Bogoliubov Laboratory of Theoretical Physics,
Joint Institute  for Nuclear Research, 141980  Dubna,
Moscow Region, Russia; and

Institute for Theoretical Physics, University of
Erlangen-N\"urnberg, Staudtstrasse 7, 91058 Erlangen, Germany}
\end{center}

\begin{abstract}

It is shown that an (anti-)self-dual homogeneous vacuum gluon field
appears in a natural way within
the problem of calculation of the QCD partition function in
the form of Euclidean functional integral
with periodic boundary conditions.
There is no violation of cluster property within this formulation,
nor are  parity,  color and rotational
symmetries broken explicitly.
The massless limit of the product
of the quark masses and condensates,
$m_f\langle\bar\psi_f\psi_f\rangle$,
is calculated to all loop orders. This quantity does not vanish and
 is proportional to the gluon condensate
appearing due to the nonzero strength of the vacuum gluon field.
We conclude that the gluon condensate
can be considered as an order parameter both for confinement and
chiral symmetry breaking.

\end{abstract}

\section{Introduction}

The physical picture of nonperturbative QCD vacuum realised with the
(anti-)self-dual homogeneous gluon field has become
prominent
since the early eighties, when Leutwyler
demonstrated the stability of this gluon configuration against local
quantum fluctuations and noticed, that this field could
be related to the problems of confinement and chiral symmetry
breaking~\cite{leutw1,leutw2}. Elizalde and Soto, and many other authors
have obtained strong evidence that this field could be a true
minimum of the QCD effective potential(see~\cite{eliz,pag} and references
therein).
Manifestations of this gluon configuration in the spectrum and
weak decays of light mesons, their excited states, heavy quarkonia and
heavy-light mesons were studied
in recent papers~\cite{efin1,efin2}. The vacuum
field under consideration produces several  qualitative
regimes for masses and decay constants which are completely consistent
with experimental data. Namely, the masses of
light pseudoscalar and vector mesons
are strongly split, orbital and radial excitations of light mesons
show Regge behaviour, the mass of heavy quarkonium tends to be equal
to sum of the masses of quarks, the heavy-light meson mass
approaches the mass of the heavy quarks, and the weak decay constant for
pseudoscalar heavy-light mesons has asymptotic behaviour
$1/\sqrt{m_Q}$. Moreover, scalar and axial mesons are absent in the
spectrum as simple $q\bar q$ states, but appear in the super-fine structure
of orbital excitations of vector mesons. Quantitatively,
the masses and decay constants of mesons from all different regions of
the spectrum
are described within ten percent inaccuracy.
These different phenomena are displayed with the minimal set
of parameters: gauge coupling constant, strength of the
vacuum field and the quark masses. It looks as if the field under
consideration produces both confinement and chiral (flavour) symmetry
breaking~\cite{efin2}.
However, there are three essential gaps that hinder
justification of this physical picture. Regular formulation of the
problem about QCD ground state, realized by the
(anti)-self-dual homogeneous field,
is missed. There is no proof that this field
minimizes the QCD effective potential.
Most of the results
concerning the field under consideration are obtained
within the one-loop approximation. In this paper, we
attempt to fill in the first and third gaps.

We  construct a representation for the Euclidean
QCD generating functional which includes the vacuum field under consideration
in a self-consistent manner. Using this representation, we investigate
the massless limit of the renormalization group invariant quantity
$m\langle\bar\psi\psi\rangle_{\cal B}$ which is
a product of the quark mass and quark condensate
in the presence of vacuum (anti-)self-dual homogeneous gluon field.
An important result of this work is the formula
\begin{equation}
\label{res1}
\sum\limits_{f=1}^{N_F}\lim_{m_f\to0}m_f
\langle\bar\psi_f\psi_f\rangle_{\cal B}=
-N_F\frac{{\cal B}^2}{\pi^2},
\end{equation}
where ${\cal B}$ is the strength of the vacuum field (gauge coupling constant
is included into ${\cal B}$).
Equation (\ref{res1}) is valid to all loop orders.

\section{Generating functional}

First of all, we need to
explain what is hidden behind the symbol
$\langle\dots\rangle_{\cal B}$. In other words,
what is the formal statement of the problem about QCD ground state,
within which this field appears in a natural and self-consistent way?

The usual statement of the problem about vacuum (phase) structure of
quantum field  systems is based on the analogy between the functional
integrals in Euclidean QFT and the partition function of
quantum statistical systems
in the infinite volume (thermodynamic) limit.
Therefore, we need an appropriate representation for
the QCD partition function in the infinite volume limit.
The most subtle point is a choice of boundary conditions for
the functional space of integration.
The standard way is to introduce a large space-time box, to impose
periodic boundary conditions on the fields in the box and then to
study the infinite volume limit. We follow just this prescription.

It should be noted, that
condition for the fields to vanish at
infinity, which is normal for QFT in the perturbative regime,
is not appropriate, since the translation invariant
fields are excluded ad hoc. There is no chance to get insight into the
critical phenomena of long range correlations.
 The instanton-like formulation of the problem --
to calculate a transition amplitude from the field configuration $A$
given at Euclidean time $\tau_1$ to another configuration $A^\prime$
at time $\tau_2$~\cite{leutw2} --
is not suitable either. In this case, the homogeneous
field comes through the boundary conditions, which results in a
hard violation of the cluster property and explicit breakdown of
rotational and color symmetries and parity.

Let us start with pure gluodynamics.
A naive representation for partition function looks like
\begin{eqnarray}
\label{zym1}
Z\sim
\int_{{\cal F}_{L,\beta}}
D{\cal A}
\exp\left\{\int_V d^4x
{\cal L}_{\rm YM}({\cal A})\right\},
\end{eqnarray}
where $V$ is a large Euclidean volume, $L$ and $\beta^{-1}=T$ are the
space box size and the temperature.
The functional space ${\cal F}_{L,\beta}$ contains gauge fields
${\cal A}_\mu$ satisfying
periodic boundary condition.
Notice, that
translation invariant fields,
in particular the (anti-)self-dual homogeneous field
\begin{eqnarray}
\label{b-field}
&&B_\mu(x)=\frac{1}{2}
nB_{\mu\nu}x_\nu, \ \ n=t_3\sin\xi+t_8\cos\xi,
\\
&&
\tilde B_{\mu\nu}=\pm B_{\mu\nu},
 \ \  B_{\mu\rho}B_{\rho\nu}=-\delta_{\mu\nu}B^2, \ \ B^2={\rm const},
\nonumber
\end{eqnarray}
belong to ${\cal F}_{L\beta}$.
In case (\ref{b-field}), an arbitrary translation
\begin{equation}
\label{trb}
B_\mu(x+\xi)=B_\mu(x)+
\partial_\mu\omega(x,\xi), \ \  \omega=x_\nu B_\nu(\xi),
\end{equation}
can be compensated by a suitable
gauge transformation.

Field configuration (\ref{b-field})  is not a dynamical variable in
the sense, that its equation of motion
does not contain any derivatives, but is just a constraint.
This field
must be integrated out if one looks for
an integral representation for partition function which corresponds to
an actual ground state of the system.
However,  this integration
should be based on resolving the constraint
which takes into account all quantum corrections
coming from the dynamical modes of the gauge fields.  The quantum constraint
can have solutions that are neither visible at the classical level nor
within the
perturbation theory. At the same time, these nontrivial solutions for
the constant fields (condensates) govern  critical phenomena in
systems with the infinite number of
degrees of freedom. One can easily illustrate
this statement by the phase transitions in the models with
$\phi^4$ and Yukawa interactions
(e.g., see~\cite{book} and references therein).

The integral over the homogeneous field can be separated in (\ref{zym1})
with a simultaneous  fixing of
a gauge of dynamical fields
by means of the Faddev-Popov trick:
\begin{eqnarray}
\label{unity2}
1=\Phi[{\cal A}]\int\limits_{\bar{\cal F}}DA
\int D\omega
\int\limits_{0}^{\infty}dB
\int\limits_{\Sigma_B}d\sigma_{B} \
\delta\left[{\cal A}-A^\omega-B^\omega\right]
\delta\left[\nabla_\mu^{ab}(B) A^a_\mu\right],
\nonumber
\end{eqnarray}
where the space $\bar{\cal F}$ does not contain the nondynamical mode
(\ref{b-field}), $B$ is the field strength,
 $\nabla_\mu(B)=\partial_\mu-iB_\mu(x)$ denotes a covariant derivative
in the adjoint representation. The coupling constant $g$ is included into
the field $B_\mu$.
The measure $d\sigma_B$  is defined as
\begin{eqnarray}
\label{sb}
\int\limits_{\Sigma_B}d\sigma_B=
\frac{1}{(4\pi)^2}\sum\limits_{\pm}
\int\limits_{0}^{2\pi}d\varphi
\int\limits_{0}^{\pi}d\theta\sin\theta
\int\limits_{0}^{2\pi}d\zeta=1.
\end{eqnarray}
Definition (\ref{sb}) corresponds to  integration over the
spatial (spherical) angles $(\varphi,\theta)$ of the field (\ref{b-field})
and angle $\xi$  which
defines its  orientation in color space (in the diagonal representation
of matrix $n$ in (\ref{b-field})).
The sign '$\pm$' corresponds to summation
of the self- and anti-self-dual configurations.
The final representation for $Z$ is then
\begin{eqnarray}
\label{zym2}
Z=\lim_{\Lambda\to\infty}{\rm R}_\Lambda N
\int\limits_{\Sigma_B}d\sigma_{B}
\int\limits_{0}^{\infty}dB
\int\limits_{\bar{\cal F}}DA
 \
\Delta_{\rm FP}[B,A]
\delta\left[\nabla(B)A\right]
\nonumber\\
\times\exp\left\{\int_V d^4x
{\cal L}_{\rm YM}(A+B)\right\},
\end{eqnarray}
where
$\Delta_{\rm FP}[B,A]$ is the Faddeev-Popov determinant for the
background gauge condition $\nabla(B)A=0$. An appropriate regularization
$R_\Lambda$ of ultraviolet divergences and renormalization prescription
are implied in Eq.~(\ref{zym2}).
By  definition, an integral over the fields $A$ gives rise to
an effective potential of the field $B_\mu$.
The background field does not affect general renormalizability of the
theory~\cite{dewitt,abbott},
and we  rewrite
$Z$ in the form
\begin{eqnarray}
\label{zym3}
Z=N'
\int\limits_{\Sigma_B}d\sigma_{B}
\int\limits_{0}^{\infty}dB
\exp\left\{-V U_{\rm eff}[B^2;g(\mu),\mu,\beta]\right\},
\end{eqnarray}
where $\mu$ is the renormalization point.
As has been mentioned, the background field
in Eq.~(\ref{zym3}) includes the coupling constant $B\equiv gB$.
In the background gauge, the composition $gB$ is RG-invariant~\cite{abbott}.
Furthermore, the effective potential is invariant
under  gauge and parity transformations, and space rotations
(this also follows from the general background field method).
Thus, we arrive at the expression
\begin{eqnarray}
Z=N'
\int\limits_{0}^{\infty}dB
\exp\left\{-V U_{\rm eff}[B^2;g(\mu),\mu,\beta]\right\}.
\nonumber
\end{eqnarray}
If the effective potential has a minimum at nonzero field strength
${\cal B}={\cal B}(g(\mu),\mu,\beta),$
then, in the infinite volume limit, the saddle-point method gives
\begin{eqnarray}
&&Z=
\exp\left\{-VF[g(\mu),\mu,\beta]\right\},\nonumber\\
&&F=U_{\rm eff}[{\cal B}^{2}(g(\mu),\mu,\beta);g(\mu),\mu,\beta]<0.
\nonumber
\end{eqnarray}
The free energy density $F$ is RG-invariant.
For zero temperature,
${\cal B}$ is nothing other than the RG-invariant combination of the running
coupling constant $g(\mu)$ and the renormalization point $\mu$, hence:
\begin{eqnarray}
\label{lq}
\lim_{\beta\to\infty}{\cal B}^{2}=C_{\cal B} \ \Lambda^4_{\rm QCD},
\ \
\lim_{\beta\to \infty}F
=-C_F\Lambda^4_{\rm QCD},
\end{eqnarray}
where $C_{\cal B}$ and $C_F$ are positive numbers, and
$$
\Lambda_{\rm QCD}^2=
\mu^2\exp\left\{\int\limits^{g(\mu)}\frac{dg}{\beta(g)}\right\}.
$$
These equations link the strength of the vacuum field with
the ``fundamental scale'' $\Lambda_{\rm QCD}$
(see also~\cite{leutw2,pag,mink}).

Now we can represent the partition function $Z$ in the form of functional
integral over the fields $A$, that does not contain the translation invariant
mode:
\begin{eqnarray}
\label{zym4}
Z=\lim_{\Lambda\to\infty}{\rm R}_\Lambda N
\int\limits_{\Sigma_{{\cal B}}}d\sigma_{{\cal B}}
\int\limits_{\bar{\cal F}}DA
 \
\Delta_{\rm FP}[{\cal B},A]
\delta\left[\nabla({\cal B})A\right]
\exp\left\{\int_V d^4x
{\cal L}_{\rm YM}(A+{\cal B})\right\}.
\end{eqnarray}
Equation (\ref{zym4}) gives the representation that we are looking for.
It is based on the strong but single assumption that the
(anti-)self-dual field corresponds to the global minimum of the QCD effective
action. Lattice calculation of the effective potential for
different translation invariant gluon fields seems to be
the most direct way to verify this assumption.
However, the (anti-)self-dual field
is a particularly interesting configuration
due to other reasons to be discussed below
(see also~\cite{leutw1,leutw2}).

On the basis of representation (\ref{zym4}),
the QCD generating functional for correlation functions
in the infinite volume and at zero temperature has to
 be defined as
\begin{eqnarray}
\label{gf1}
&&Z_{\cal B}[J,\eta,\bar\eta]=
\lim_{\Lambda\to\infty}{\rm R}_\Lambda{\cal N}_{\cal B}
\int\limits_{\Sigma_{\cal B}}d\sigma_{\cal B}
\int\limits_{\bar{\cal F}}D\mu_{A}(A,{\cal B})
\int\limits_{\cal G}\prod\limits_{f}D\psi_f D\bar\psi_f
\nonumber\\
&&\exp\left\{\int d^4x
\bar\psi_f(x)
\left[i\hat\nabla-m_f+g\hat A\right]\psi_f(x)
+i\int d^4x\left[JA+\bar\eta_f\psi_f+\bar\psi_f\eta_f\right]\right\},
\nonumber\\
&&D\mu_{A}(A,{\cal B})=DA
\Delta_{\rm FP}[{\cal B},A]
\delta\left[\nabla({\cal B})A\right]
\exp\left\{\int d^4x
{\cal L}_{\rm YM}[A+{\cal B}]\right\},
\\
&& \hat A=\gamma_\mu A_\mu, \ \
\hat\nabla=\gamma_\mu\nabla_\mu, \ \
\nabla_\mu=\partial_\mu-i{\cal B}_\mu.
\nonumber
\end{eqnarray}
The constant ${\cal N}_{\cal B}$ provides the standard normalization
$Z_{\cal B}[0,0,0]=1$.
The functional space $\bar{\cal F}$
contains the gauge fields vanishing at infinity. The change
of boundary conditions (vanishing fields instead of periodic ones)
is unimportant
for physics, since quantum
fluctuations $A$ does not contain translation invariant modes.
We have also added massive quarks.
The fermionic
functional integral spans the
Grassmann algebra ${\cal G}$ of square integrable
fields.
To be more precise, we will
 define this integral  via a
decomposition of the fields $\bar\psi$ and $\psi$
over the eigenmodes
$\psi_n$ of the Dirac operator in the presence of vacuum
gluon field ${\cal B}$ (an anti-hermitian representation
for the $\gamma$-matrices in Euclidean space is used)
\begin{eqnarray}
\label{Dir1}
-i\hat\nabla\psi_n(x)
=i\lambda_n\psi_n(x).
\end{eqnarray}
As a matter of fact, this definition of the fermionic
integral implies, that the ground state of the system is governed by
the vacuum field ${\cal B}$, and the interaction $\bar\psi\hat A\psi$
 of quarks with
the quantum gauge field $A$ has to be treated as
perturbation.
Now, let us seek insight into the properties of representation (\ref{gf1}).

\section{Parametrization, cluster property, symmetries}

The generating functional (\ref{gf1}) contains the intrinsic
dimensionful quantity ${\cal B}$ (see also (\ref{lq})),
which provides the natural reference scale for running quark masses
$\bar m_f(\mu)$ and
gauge coupling constant $\bar\alpha_s(\mu)$. Therefore, the
 strength of the vacuum field ${\cal B}$,
the quark masses and coupling constant at
the scale $\mu=\sqrt{\cal B}$ can be considered as the physical
(intrinsic) parameters of QCD in the representation (\ref{gf1}).
Values of the parameters can be extracted from the analysis of
hadron spectrum (e.g., see \cite{efin2}).

Correlation functions for the local or nonlocal
operators ${\cal O}_j[A,\psi,\bar\psi]$ defined in the standard way
($A\in\bar{\cal F}$~!)
\begin{eqnarray}
\label{corrym1}
&&\langle {\cal O}_1[A,\psi,\bar\psi]\dots
{\cal O}_n[A,\psi,\bar\psi]\rangle_{{\cal B}}=
\nonumber\\
&& \ \ \ \ \
\left( {\cal O}_1\left[\frac{\delta}{i\delta J},
\frac{\delta}{i\delta\bar\eta},\frac{\delta}{i\delta\eta}\right]
\dots
{\cal O}_n\left[\frac{\delta}{i\delta J},
\frac{\delta}{i\delta\bar\eta},\frac{\delta}{i\delta\eta}\right]
Z_{\cal B}[J,\eta,\bar\eta]\right)_{J=\eta=\bar\eta=0}
\end{eqnarray}
ensure the cluster property.
Due to the integration over the angular variables $\Sigma_{\cal B}$
and summation of the self- and anti-self-dual configurations,
the parity, rotational and color symmetries
are not  broken explicitly.
The correlators depend only on  ${\cal B}^2$ and have normal
transformation properties. At the same time, violation of the symmetries
is seen in the integrand of Eq.~(\ref{gf1}), which is an indication
of spontaneous breaking of the above-mentioned symmetries.
Thus, we meet very unusual mechanism of SSB due to the condensation
of the vector bosons.
The order parameter for the nonperturbative phase
is obvious. This is the lowest nonvanishing gluon correlator
(gluon condensate), defined as
\begin{eqnarray}
\label{corrym3}
\langle \left[\partial_\nu{\cal A}_\mu^a(x)-
\partial_\mu{\cal A}_\nu^a(x)\right]^2\rangle
= 4{\cal B}^{2}+({\rm pert. corr.}).
\end{eqnarray}
Here ${\cal A}=(A+B)\in {\cal F}$, and $\langle\dots\rangle$ denotes an
averaging by means of Eq.~(\ref{zym2}).
However, any of the correlators
\begin{eqnarray}
\label{corrym2}
\langle {\cal O}_1[{\cal A}]\cdot...\cdot{\cal O}_n[{\cal A}]\rangle=
\int_{\Sigma_{\cal B}}d\sigma_{\cal B}
{\cal O}_1[{\cal B}]\cdot...\cdot{\cal O}_n[{\cal B}] +...,
\end{eqnarray}
contains a constant part and can be taken as order parameter.

\section{Chiral symmetries}

Spontaneous violation of parity should influence
the chiral symmetries $U_A(1)$ and $SU_L(N_F)\times SU_R(N_F)$.
Due to summing the self- and anti-self-dual configurations
in (\ref{gf1}), the vacuum expectation value
of a  pseudo-tensor operator
is identically equal to zero. In particular,
$U_A(1)$  symmetry is not broken in the sense that
$$
\langle\partial_\mu\bar\psi_f(x)\gamma_5\gamma_\mu
\psi(x)\rangle_{\cal B}\equiv 0.
$$
An explicit
violation of parity as in the instanton $\theta$-vacuum is needed.

To study the flavour chiral symmetry $SU_L(N_F)\times SU_R(N_F)$
let us
consider the massless limit of composition of
the quark masses and quark condensates:
\begin{eqnarray}
\label{qc1}
&&\sum\limits_{f=1}^{N_F} \
m_f\langle \bar\psi_f(x) \psi_f(x)\rangle_{{\cal B}}=
\sum\limits_{f=1}^{N_F}
m_f\left[\frac{\delta}{i\delta\eta_f(x)}\frac{\delta}{i\delta\bar\eta_f(x)}
Z_{\cal B}[\eta,\bar\eta,J]\right]_{\bar\eta=\eta=J=0}.
\end{eqnarray}
The nontrivial point in the calculation of this quantity consists
in the following.
In the massless limit $m_f\to0$,
the  divergent contributions $O(m_f^{-k})$, with $k$
being some positive integer, can appear potentially at any loop
order. This means, that the perturbation decomposition can fail
in the massless limit. In this case, the
divergent terms have to be summed to all loop orders.
This infrared problem comes
>from the zero modes $\psi_0$ of Eq.~(\ref{Dir1}) with
$\lambda_0=0$ existing due to an (anti-)self-duality of the vacuum field.
It should be noted, that this problem arises both for
the homogeneous and instanton fields. However, the crucial difference
between the homogeneous field and the instanton $\theta$-vacuum consists
in the normalization of the generating functional.
Unlike the instanton case
(e.g., see~\cite{coleman,car}), the normalization constant
${\cal N}_{\cal B}$ in Eq.~(\ref{gf1}) corresponds to the
nonperturbative vacuum and  contains the  contribution of the
fermion zero modes. Therefore, in the massless limit, no problem arises
with the fermion determinant coming from the
integral in (\ref{gf1}); it is cancelled by the normalization constant
${\cal N}_{\cal B}$.

Now we will show that a
singularity $1/m$ exists in the lowest one-loop
diagram for the quark condensate but does not appear in  higher
orders. This allows one to calculate the massless
limit of Eq.~(\ref{qc1}) explicitly and to prove relation
(\ref{res1}).
The most direct way consists in the following.

First of all, notice that Eq.~(\ref{qc1}) can be rewritten in the
equivalent form
\begin{eqnarray}
\label{qc2}
&&\sum\limits_{f=1}^{N_F} \
m_f\langle \bar\psi_f(x) \psi_f(x)\rangle_{{\cal B}}=
-{\cal Z}^{-1}_{\cal B}(m)\sum\limits_{f=1}^{N_F}\lim_{V\to\infty}
V^{-1}
m_f\frac{d}{d m_f}
{\cal Z}_{\cal B}(m),\\
&&{\cal Z}_{\cal B}(m)=
\lim_{\Lambda\to\infty}{\rm R}_\Lambda{\cal N}_{\cal B}(\mu)
\int\limits_{\Sigma_{\cal B}}d\sigma_{\cal B}
\int\limits_{\bar{\cal F}}D\mu_{A}(A,{\cal B})
\int\limits_{\cal G}\prod\limits_{f}D\psi_f D\bar\psi_f
\\
&&\exp\left\{\int d^4x
\bar\psi_f(x)
\left[i\hat\nabla-m_f+g\hat A\right]\psi_f(x)
\right\},
\nonumber
\end{eqnarray}
where the normalization constant ${\cal N}_{\cal B}(\mu)$ is taken so that
${\cal Z}_{\cal B}(\mu)=1$ for some $\mu\not=0$.
The LHS of Eq.~(\ref{qc2})
does not depend on ${\cal N}_{\cal B}(\mu)$, but this normalization
provides us with an appropriately defined integral under the derivative.

Consider for a moment the one-flavour case. An extension to
$N_F>1$ is straightforward.
Formal integration over the quark field in the partition function
gives
\begin{eqnarray}
\label{gf11}
Z_{\cal B}(m)=
\lim_{\Lambda\to\infty}{\rm R}_\Lambda {\cal N}_{\cal B}(\mu)
\int\limits_{\Sigma_{\cal B}}d\sigma_{\cal B}
\int\limits_{\bar{\cal F}}D\mu_{A}(A,{\cal B})
{\rm det}\left[-i\hat\nabla+m-g\hat A\right].
\end{eqnarray}
Our definition of the fermion
integral via the eigenmodes of Dirac operator means the
determinant in (\ref{gf11}) and its derivative in (\ref{qc2})
are defined as
\begin{eqnarray}
\label{detdef}
{\rm det}\left[-i\hat\nabla+m-g\hat A\right]=
{\rm det}\left[-i\hat\nabla+m\right]
{\rm det}\left[1-g\hat AS\right],
\\
\label{detder}
\frac{d}{dm}{\rm det}\left[-i\hat\nabla+m-g\hat A\right]=
{\rm det}\left[-i\hat\nabla+m\right]
{\rm det}\left[1-g\hat AS\right]\times
\nonumber\\
\left[\tilde {\rm Tr}S+
\frac{d}{dm}\tilde {\rm Tr}\ln(1-g\hat AS)\right],
\end{eqnarray}
where the trace $\tilde {\rm Tr}$ includes  the space-time integration,
and the quark propagator $S(x,y)$ satisfies the equation
\begin{eqnarray}
\label{s}
\left(i\hat\nabla_x-m\right)S(x,y)=-\delta(x-y).
\end{eqnarray}
The term $\tilde {\rm Tr}S$ in (\ref{detder}) is the lowest order
contribution to the quark
condensate. Higher perturbation corrections come from the quark
loops contained in the logarithmic term in (\ref{detder}).
The decisive point is that these quark loops are regular in the
massless limit, while the lowest term is singular:
\begin{eqnarray}
\label{lim1}
\lim_{m\to0}\tilde {\rm Tr}\ln(1-g\hat AS)\sim 1+O(m), \ \
\lim_{m\to 0}\tilde {\rm Tr}S\sim \frac{1}{m}+O(1).
\end{eqnarray}
It is notable, that, in another context,
 a regularity of the simplest two-gluon loop
was demonstrated by Flory~\cite{flory}.

Using the standard representation for Green's function
in terms of the matrix elements of the projection
operators ${\cal P}_n$
$$
S(x,y)=\sum\limits_{n=0}^{\infty}
\frac{{\cal P}_n(x,y)}{m+i\lambda_n},
$$
one can separate the contribution of the zero eigenmodes
and normal modes to the propagator
\begin{eqnarray}
\label{qpro01}
&&S(x,y)=S_0(x,y)+S^\prime(x,y),
\nonumber\\
&& S_0(x,y)={\cal P}_0(x,y)/m, \\
\label{qpropn1}
&&S^\prime(x,y)=\stackrel{\rightarrow}{i\hat\nabla_x}
\Delta(x,y)P_{\pm}+
\Delta(x,y)\stackrel{\leftarrow}{i\hat\nabla_y}P_{\mp}
+O(m),\\
&&\stackrel{\rightarrow}{\nabla}=\stackrel{\rightarrow}{\partial}
-i{\cal B}, \ \
\stackrel{\leftarrow}{\nabla}=\stackrel{\leftarrow}{\partial}
+i{\cal B}.
\nonumber
\end{eqnarray}
Here ${\cal P}_0$ is the projector onto the zero mode subspace
\begin{eqnarray}
\label{p0}
&&\int d^4z{\cal P}_0(x,z){\cal P}_0(z,y)={\cal P}_0(x,y),
\nonumber\\
&&{\cal P}_0=\frac{n^2{\cal B}^2}{4\pi^2}f(x,y)P_{\pm},
\\
\label{f0}
&&f(x,y)=\exp\left\{
-\frac{1}{4}\sqrt{n^2}{\cal B}(x-y)^2+
\frac{i}{2} n x_\mu{\cal B}_{\mu\nu}y_\nu\right\},
\end{eqnarray}
$\Delta(x,y)=f(x,y)/4\pi^2(x-y)^2$ is the scalar massless propagator
in the background field (\ref{b-field}), $n$ is a diagonal matrix
(see Eq.~(\ref{b-field})), and $P_\pm=(1\pm\gamma_5)/2$.

Representation (\ref{p0}) is defined by the general square
integrable  solution
\begin{eqnarray}
\label{zm1}
&&\psi_0(x,x_0)=\frac{ (n^2{\cal B}^2)^{3/4}}{4\pi^2}
i\hat\nabla_x uf(x,x_0)
\end{eqnarray}
of the equation (\ref{Dir1}) with $\lambda_0=0$. Details of calculation
of $\psi_0$ and ${\cal P}_0$ can be found in Appendix.
The space-time point $x_0$ describes a
position of the fermion ``pseudoparticle'' $\psi_0(x,x_0)$.
We see that the zero eigenvalue is infinitely degenerate which is a
manifestation of the above-mentioned invariance of the vacuum field
under
translations and simultaneous gauge transformations (see (\ref{trb})).
This feature, as well as the functional form of fermion zero mode
(\ref{f0}), (\ref{zm1}), is very similar to the properties of Leutwyler's
chromons~\cite{leutw1,leutw2}
which are gluon zero modes in the same background field.

The spinor $u$ in Eq.~(\ref{zm1}) is an eigenvector of the $\gamma_5$-matrix
$\gamma_5\psi_0=\pm\psi_0$,
which is the well-known~\cite{coleman,brown1,brown2} property of
zero modes to be right-handed in a self-dual field and
left-handed in an anti-self-dual field. As a result, the projector
$$
{\cal P}_0(x,y)=\int d^4x_0\psi_0(x,x_0)\psi_0^\dagger(y,x_0)
$$
contains the chiral projection matrix $P_{\pm}$ (see (\ref{p0})).

Representation (\ref{qpropn1}) for the normal mode propagator
was obtained by Brown {\it et al}~\cite{brown1}
for an arbitrary (anti-)self-dual background field (see also~\cite{brown2}).

Now we can return to Eq.~(\ref{detder}) and represent
the logarithmic term in the form
\begin{equation}
\label{loops}
\tilde {\rm Tr}\ln(1-g\hat AS)=
\sum\limits_{k=1}^{\infty}\frac{(-g)^k}{k}
\tilde {\rm Tr}\left[\hat A(S_0+S^\prime)\right]^k.
\end{equation}
Then one makes use of Eqs.~(\ref{qpro01}) and (\ref{p0}) to notice
that
\begin{equation}
\label{decoupl}
S_0(x,y)\gamma_\mu A_\mu(y)S_0(y,z)\equiv0
\end{equation}
due to the projectors $P_{\pm}$ in $S_0$. Therefore,
all the terms in Eq.~(\ref{loops}), which contain the block
$S_0\gamma_\mu A_\mu S_0$, vanish.
Furthermore, any two propagators
$S_0\sim P_{\pm}$ in the rest of terms of Eq.~(\ref{loops}) are separated
by an odd number of vertices $\gamma A$ and propagators
$S^\prime\sim (\gamma+O(m))$. Hence, the terms in (\ref{loops})
with nonzero trace of $\gamma$-matrices contain at least one quark mass
$m$ in the numerator for each $m$ in the denominator. In other words,
there is always an odd number of $\gamma$-matrices between two
chiral projectors $P_{\pm}$ in the singular terms, and
the quark loops are finite in the limit $m\to 0$, as is pointed out
in Eq.~(\ref{lim1}).

Finally, taking into account equations (\ref{qc2}),
(\ref{gf11}), (\ref{detder}) and (\ref{qpro01})-(\ref{f0}),
we arrive at
\begin{equation}
\label{onefl}
\lim_{m\to0}m\langle\bar\psi\psi\rangle_{\cal B}=-\lim_{m\to0}
m\frac{1}{V}\tilde {\rm Tr}S=-V^{-1}\int_V d^4x {\rm Tr}
{\cal P}_0(x,x)=
-\frac{{\cal B}^2}{\pi^2}
\end{equation}
for one flavour.  For several flavours one gets formula (\ref{res1}).
Thus, the gluon condensate (\ref{corrym3})
can be considered as an order parameter
for the flavour chiral symmetry breaking.

The nonzero massless limit of
$m\langle\bar\psi\psi\rangle_{\cal B}$ indicates a non-Goldstone
mechanism of symmetry breaking. From our point
of view, breakdown of the chiral symmetry appears here as a secondary effect
of spontaneous violation of parity, which is a discrete symmetry.
Since zero modes (\ref{zm1}) are left-handed in the anti-self-dual field
and right-handed in the self-dual field, hence
in both terms ($\pm$) of generating functional (\ref{gf1})
the chiral
group is reduced to one of the flavour subgroups
$$
SU_L(N_F)\times SU_R(N_F) \longrightarrow SU_L(N_F) \ \
({\rm or} \ SU_R(N_F))
$$
for the zero mode
component of the fermion fields
$$
\chi_0(x)=\int d^4z q_0(z)\psi_0(x,z), \ \
\bar\chi_0(x)=\int d^4z \bar q_0(z)\psi^\dagger_0(x,z),
$$
where $(q_0,\bar q_0)$ are the basic elements  of the zero mode
subspace of the Grassmann
algebra ${\cal G}$ in Eq.~(\ref{gf1}).
As has been mentioned,
due to the translation invariance of the vacuum field,
there is a continuum of fermion zero modes,
and their condensation in the
infinite volume produces a very strong effect.
Consequences of this effect in meson phenomenology
are discussed in~\cite{efin2}.

\section{Confinement}

Now we will comment briefly on the quark confinement produced
by the field under consideration. Fourier transform of
the two-point quark correlator
defined by Eqs.~(\ref{gf1}) and (\ref{corrym1})
\begin{eqnarray}
\langle \psi_f(x)\bar\psi_f(y)\rangle_{\cal B}=
\int_{\Sigma_{\cal B}}d\sigma_{\cal B}\sum\limits_{\pm}S_f(x,y)\ +
\ ({\rm pert. corr.}),
\end{eqnarray}
with $S$ being the solution to Eq.~(\ref{s}), is an entire analytical function
in the complex momentum plane (for explicit form of $S$
see~\cite{efin1,efin2,efiv}). This means that there are no poles corresponding
to free quarks. The other side of this peculiarity is that
the Dirac equation for massive quarks in the presence of
the background field~(\ref{b-field})
$$
\left(i\hat\nabla_x-m_f\right)\psi(x)=0
$$
has only the trivial solution $\psi\equiv0$.
Therefore, one has no appropriate field  to construct asymptotic free
states for quarks.
In this sense, the quarks cannot exist as  free particles but can
propagate as virtual objects. The characteristic scale of propagation
of these quark ``virtons'' is determined by the strength ${\cal B}$
of the vacuum gluon field. This situation can be seen as
the quark confinement, for which
gluon condensate ${\cal B}^2$  plays the role of
an order parameter.
Meantime,
nothing preserves these virtons to form
a colorless composite particle by means of gluon exchange~\cite{efin1,efin2}.
A colorless bound state does not feel the confining vacuum field and can
be observable. A mathematically consistent treatment of this physical concept,
especially a reasonable solution of the bound state problem in
terms of composite fields, requires an application of the methods
of nonlocal quantum field theory~\cite{efin1,efin2,efiv,mono}.

In conclusion we would like to mention the ``flaws''
in this picture.
The problem about the minimum of the effective potential is not solved.
Besides confined modes of the
gluon field (in the same sense as for the quarks),
 free gluons appear to be allowed.
At first sight,
the gluons, longitudinal in the color space with respect to
the vacuum field, seem to be not confined~\cite{leutw2,efin1}.
Solution of the $U_A(1)$ problem, which is missed
in our consideration, can come from the investigation of the
local instanton-like (anti-)self-dual deformations of the homogeneous
background field - chromons~\cite{leutw2,mink}. In the meantime, these
``flaws'' are problems for further consideration rather than
 reasons to reject the whole physical concept.

\section{Appendix}

Here we will find the explicit form of the solution $\psi_0$
to Eq.~(\ref{Dir1}) corresponding to the zero eigenvalue:
\begin{eqnarray}
\label{ccsb1}
\gamma_\mu\nabla_\mu\psi_0=0,
\end{eqnarray}
where the background field can be taken in the form
\begin{eqnarray}
&&\nabla_\mu=\partial_\mu-iB_\mu, \ \ B_\mu=\frac{1}{2}nB_{\mu\nu}x_\nu,
\ \ n=t_3\sin\zeta+t_8\cos\zeta.
\nonumber\\
&&B_{12}=B, \ \ B_{34}=\epsilon B, \ \ \epsilon=\pm 1,
\nonumber\\
&&B_{\mu\rho}B_{\rho\nu}=-B^2\delta_{\mu\nu},
\ \ \tilde B_{\mu\nu}=\epsilon B_{\mu\nu}.
\nonumber
\end{eqnarray}
Let us solve the eigenvalue problem for the squared Dirac operator
\begin{eqnarray}
\label{ccsb2}
\left(-\nabla^2+\frac{n}{2}\sigma_{\mu\nu}B_{\mu\nu}\right)\phi=
\xi\phi.
\end{eqnarray}
According to (\ref{ccsb1}), the zero mode $\psi_0$
has the form
\begin{eqnarray}
\psi_0=i\gamma_\mu\nabla_\mu\phi_0,
\nonumber
\end{eqnarray}
where $\phi_0$ is the zero mode solution ($\xi=0$) of Eq.~(\ref{ccsb2}).
Because of the relations
($'+'$ -- self-dual field,
$'-'$ -- anti-self-dual field)
$$
\sigma_{\mu\nu}B_{\mu\nu}P_\pm=0, \ \
[\sigma_{\mu\nu}B_{\mu\nu},\gamma_5]=0,
$$
solution $\phi$ of Eq.~(\ref{ccsb2}) is a (right-)left-handed spinor
for the (anti-)self-dual field, and it can be represented in the form
\begin{equation}
\label{ccsb3}
\phi(x)=u^{(\mp)}_sf(x),  \ \ \
\gamma_5 u_s^{(\pm)}=\pm u_s^{(\pm)}, \ \ \ s=1,2,
\end{equation}
where $f(x)$ is a scalar function.
In the Weil representation
\begin{eqnarray}
&&u_1^{(-)\dagger}=(1,0,0,0), \ \ \
u_2^{(-)\dagger}=(0,1,0,0), \nonumber\\
&&u_1^{(+)\dagger}=(0,0,1,0), \ \ \
u_2^{(+)\dagger}=(0,0,0,1), \nonumber\\
&&
P_{\pm}=\frac{1}{2}(1\pm\gamma_5)=\sum\limits_{s=1,2}P_\pm^{(s)}, \ \
P_\pm^{(s)}=u_s^{(\pm)}u_s^{(\pm)\dagger},
\nonumber\\
&&
P_-^{(1)}={\rm diag}(1,0,0,0), \ \
P_-^{(2)}={\rm diag}(0,1,0,0), \nonumber\\
&&P_+^{(1)}={\rm diag}(0,0,1,0), \ \
P_+^{(2)}={\rm diag}(0,0,0,1). \nonumber \\
\label{ccsb5}
&&\sigma_{\mu\nu}B_{\mu\nu}=4B\left[
\frac{1+\epsilon}{2}P_-^{(1)}-
\frac{1+\epsilon}{2}P_-^{(2)}+
\frac{1-\epsilon}{2}P_+^{(1)}-
\frac{1-\epsilon}{2}P_+^{(2)}
\right],
\nonumber\\
&&
\sigma_{\mu\nu}B_{\mu\nu}u_s^{(\pm)}=-(-1)^s\frac{1\mp\epsilon}{2}
u_s^{(\pm)}.
\end{eqnarray}
The following useful relations take place
\begin{eqnarray}
\label{ccsb4}
&&
P_\pm^{(s)}P_\pm^{(s^\prime)}=\delta_{ss^\prime}P_\pm^{(s)}, \ \ \
P_\pm^{(s)}P_\mp^{(s^\prime)}=0 \ \ \
(s=1,2,\ s^\prime=1,2),
\nonumber\\
&&P_\pm^{(s)}\gamma_\mu=\gamma_\mu P_\mp^{(s)}, \ \ \
P_\mp^{(s)}\gamma_\mu=\gamma_\mu P_\pm^{(s)}, \ \ \
\mu=1,2,
\nonumber\\
&&
P_\pm^{(s)}\gamma_\mu=\gamma_\mu P_\mp^{(s^\prime)}, \ \ \
P_\mp^{(s)}\gamma_\mu=\gamma_\mu P_\pm^{(s^\prime)}, \ \ \
\mu=3,4, \ \ s\not=s^\prime.
\end{eqnarray}
Taking into account Eqs.~(\ref{ccsb2})--(\ref{ccsb5}), one obtains
\begin{equation}
\label{ccsb6}
\left[-\nabla_x^2-(-1)^s(1\mp\epsilon)nB\right]
f(x)=\xi^\pm_{s} f(x).
\end{equation}
It should be stressed here, that for an
arbitrary translation $x\to x-x_0$ the differential operator
in the LHS of Eq.~(\ref{ccsb6})  transforms as
[$(xBx_0)\equiv x^\mu B_{\mu\nu} x^\nu_0$]
\begin{eqnarray}
\left[-\nabla_x^2-(-1)^s(1\mp\epsilon)nB\right]\rightarrow
e^{-\frac{in}{2}(xBx_0)}
\left[-\nabla_x^2-(-1)^s(1\mp\epsilon)nB\right]
e^{\frac{in}{2}(xBx_0)},
\nonumber
\end{eqnarray}
which is a result of the following transformation
property of the background field
\begin{eqnarray}
nB_{\mu\nu}x^\nu\rightarrow
nB_{\mu\nu}(x^\nu-x_0^\nu)=
e^{-\frac{in}{2}(xBx_0)}nB_{\mu\nu}x^\nu e^{\frac{in}{2}(xBx_0)}
-\frac{n}{2}\partial_\mu(xBx_0).
\end{eqnarray}
This means that each  eigenvalue $\xi^\pm_s$ is infinitely degenerate,
since for any $x_0$ the function
\begin{eqnarray}
\label{ccsb6-1}
F(x,x_0)=e^{\frac{in}{2}(xBx_0)}f(x-x_0)
\end{eqnarray}
is the eigenfunction with the eigenvalue $\xi^\pm_s$.

Equation (\ref{ccsb6}) shows that $f(x)$ is an eigenfunction of the
operator
$$
-\nabla^2=\frac{\sqrt{n^2}B}{2}
\left[
i\frac{\partial}{\partial\eta}+b(\eta)
\right]^2,
$$
where we have denoted
$$
\eta=\sqrt{\frac{B\sqrt{n^2}}{2}} x  ,  \ \
b_{\mu\nu}=\frac{n}{\sqrt{n^2}}\frac{B_{\mu\nu}}{B} \ \
\left(b_{\mu\rho}b_{\rho\nu}=-\delta_{\mu\nu} \right).
$$
Let us introduce the projection matrix
$$
Q_{\mu\nu}^{\pm}=\frac{1}{2}
\left[\delta_{\mu\nu}\pm ib_{\mu\nu}\right],
$$
$$
Q^++Q^-=I, \ \
(Q^\pm)^2=Q^\pm, \ \
Q^\pm Q^\mp=0, \ \ (Q^\pm)^{\rm T}=Q^\mp, \ \ bQ^{\pm}=\mp iQ^\pm,
$$
then the following relations take place
$$
-\nabla^2=\frac{\sqrt{n^2}B}{2}(i\partial_\mu-\eta_\rho b_{\rho\mu})
(Q^++Q^-)_{\mu\nu}(i\partial_\nu+b_{\nu\sigma}\eta_\sigma),
$$
$$
-\nabla^2=\frac{\sqrt{n^2}B}{2}(\eta+\partial)Q^+(\eta-\partial)+
\frac{\sqrt{n^2}B}{2}(\eta-\partial)Q^-(\eta+\partial).
$$
Using these formulas one can get
\begin{eqnarray}
&&-\nabla^2=2B\sqrt{n^2}(a^+Q^-a+2B\sqrt{n^2}),
\nonumber\\
&&a=\frac{1}{\sqrt{2}}(\eta+\partial), \ \
a^+=\frac{1}{\sqrt{2}}(\eta-\partial), \ \
[a_\alpha,a^+_\beta]=\delta_{\alpha\beta},
\nonumber\\
&&a^+Q^-a=(Q^+a^+)_\alpha(Q^-a)_\alpha.
\nonumber
\end{eqnarray}
Thus, we arrive at the harmonic oscillator algebra. Therefore,
equation (\ref{ccsb7}) and its eigenfunctions and eigenvalues can
be written in the form
\begin{eqnarray}
\label{ccsb8}
&&
\sqrt{n^2}B
\left[2a^+Q^-a+2
-(-1)^s(1\mp\epsilon){\rm sign}(n)
\right]
f_k=\xi^\pm_{s,k} f_k,
\\
&&\xi^\pm_{s,k_1k_2k_3k_4}=2\sqrt{n^2}B[k_1+k_2+k_3+k_4+1
-(-1)^s\frac{(1\mp\epsilon)}{2}{\rm sign}(n)],
\nonumber\\
&&f_{k_1k_2k_3k_4}=
[Q^+a^+]^{k_1}_1
[Q^+a^+]^{k_2}_2
[Q^+a^+]^{k_3}_3
[Q^+a^+]^{k_4}_4
f_0(\eta).
\nonumber\\
\label{ccsb7}
&&a_\alpha f_0(\eta)=0, \ \
f_0=\exp\left\{-\frac{1}{2}\eta^2\right\}=
\exp\left\{-\frac{1}{4}\sqrt{n^2}Bx^2\right\},
\\
&&(a^+Q^-a)f_{k_1k_2k_3k_4}(\eta)
=
(k_1+k_2+k_3+k_4)f_{k_1k_2k_3k_4}(\eta).
\nonumber
\end{eqnarray}
The sign ``$\pm$'' in (\ref{ccsb8}) relates to the left- ($+$) and
right-handed ($-$) spinors $u^{(\mp)}$ and $\epsilon=\pm1$ for the
self-dual ($+$) and anti-self-dual ($-$) field.

We are looking for the zero mode solutions
$\psi_0=i\gamma_\mu\nabla_\mu\phi_0$ of (\ref{ccsb1})
corresponding to $\xi^\pm_{s,0000}=0$ in Eq.~(\ref{ccsb8}) , i.e.,
we have to satisfy the condition
$$
(-1)^s(1\mp\epsilon){\rm sign}(n)=2,
$$
$$
s=1,2, \ \ \epsilon=-1,1, \ \ {\rm sign}(n)=-1,1.
$$
According to Eqs.~(\ref{ccsb3}), (\ref{ccsb5}), and (\ref{ccsb6-1}),
(\ref{ccsb7}),
the zero mode solution looks as
\begin{eqnarray}
\label{ccsb9}
&&\psi_0(x,x_0)=\frac{(n^2B^2)^{3/4}}{4\pi^2}
i\gamma_\mu\nabla_\mu(x) u f(x,x_0)
\\
&&f(x,x_0)=\exp\left\{-\frac{1}{4}\sqrt{n^2}B(x-x_0)^2+
\frac{i}{2}nx_\mu B_{\mu\nu}x_0^\nu\right\}
\nonumber\\
&&u= u_2^{(\mp)}
\ \ (\epsilon=\pm1, \ {\rm sign}(n)=1),
\ \ \
u= u_1^{(\mp)}
\ \ (\epsilon=\pm1, \ {\rm sign}(n)=-1).
\nonumber
\end{eqnarray}
The normalization constant is chosen to provide the proper normalization
of the projector onto the zero mode subspace
${\cal P}_0$ to be calculated below.

Relations (\ref{ccsb4}) are invariant under the change
$\pm\leftrightarrow\mp$ or $s\leftrightarrow s^\prime$, and
the zero mode projection operator can be constructed by the
procedure being common for all the possibilities
contained in (\ref{ccsb9}). We consider the case ${\rm sign}(n)=1$.
The projector operator is defined by the formula
\begin{eqnarray}
&&{\cal P}_0=\int d^4x_0\psi(x,x_0)\psi^\dagger(y,x_0)
\nonumber\\
&&=\frac{(n^2B^2)^{3/2}}{4\pi^2}\int d^4x_0i\gamma_\mu
\stackrel{\rightarrow}{\nabla}_\mu(x)
f(x,x_0)P^{(2)}_\mp f^*(y,x_0)
i\gamma_\nu\stackrel{\leftarrow}{\nabla}_\nu(y)
\nonumber   \\
&&\stackrel{\rightarrow}{\nabla}_\mu(x)=
\stackrel{\rightarrow}{\partial}_\mu-iB_\mu(x), \ \
\stackrel{\leftarrow}{\nabla}_\nu(y)=
\stackrel{\leftarrow}{\partial}_\nu-iB_\nu(y).
\end{eqnarray}
One can check that
\begin{eqnarray}
&&
\int d^4x_0f(x,x_0)f^*(y,x_0)=\frac{4\pi^2}{n^2B^2}f(x,y),
\nonumber\\
&&
\stackrel{\rightarrow}{\nabla}_\mu(x)f(x,y)=-\frac{1}{2}
\left(\sqrt{n^2}B(x-y)_\mu+inB_\mu(x-y)\right)f(x,y),
\nonumber\\
&&i\stackrel{\rightarrow}{\nabla}_\mu(x)f(x,y)
i\stackrel{\leftarrow}{\nabla}_\nu(y)=
-\frac{\sqrt{n^2}}{2}B\left[Q_{\mu\nu}^+
+ Q^+_{\mu\sigma}(x-y)_\sigma Q^+_{\mu\rho}(x-y)_\rho\right]
\gamma_\mu P^{(2)}_{\mp}\gamma_\nu,
\nonumber
\end{eqnarray}
Using Eqs.~(\ref{ccsb4}), we get
\begin{eqnarray}
&&{\cal P}_0(x,y)=\frac{n^2B^2}{4\pi^2}f(x,y)P_{\pm},
\nonumber\\
&&\int d^4z{\cal P}_0(x,z){\cal P}_0(z,y)={\cal P}_0(x,y),
\nonumber\\
&&\tilde{\rm Tr}{\cal P}_0=V\frac{B^2}{\pi^2}.
\nonumber
\end{eqnarray}

\centerline{\bf\large Acknowledgments}

We would like to thank F.~Lenz for numerous stimulating questions and
valuable comments.
We are also indebted to A.~Kalloniatis for critical reading the
manuscript and useful remarks. Fruitful
discussions with S.~Bastrukov, Ja.~Burdanov, A.~Bel'kov, M.~Thies,
L.~von~Smekal, S.~Solunin
and O.~Teriaev are gratefully acknowledged. S.N. thanks his colleagues from
the Institute of Theoretical Physics, University of
Erlangen-N\"urnberg, for kind hospitality. This work was supported by BMBF.
G.E. was supported in part  by the Russian Foundation for Basic Research
under grant No.~96-02-17435-a.

\end{document}